\documentclass[11pt]{article}
\usepackage[colorlinks=true]{hyperref}
\hypersetup{colorlinks=true,linkcolor=teal,citecolor=blue,urlcolor=blue}
\usepackage{graphicx}
\usepackage{amsmath}
\usepackage{amssymb}
\usepackage{latexsym}
\usepackage{color}
\usepackage{subfigure}
\usepackage{pst-all}
\usepackage{pstricks,pst-node,pst-text,pst-3d}
\usepackage{ifpdf}
\usepackage{multimedia}
\usepackage{lmodern}
\usepackage{cite}
\usepackage{sidecap}
\usepackage{dsfont}
\usepackage{bbm}
\usepackage{lipsum}
\usepackage{url}
\usepackage{textcomp}
\usepackage{bbold}
\usepackage{epstopdf}
\usepackage{epsfig}
\usepackage{mathtools}
\usepackage{nccmath}
 \usepackage[normalem]{ulem}
\usepackage{ifpdf}
\ifpdf
\else
  \usepackage{pstricks}
\fi
  \hypersetup{pdfstartview=XYZ}
\newcommand{\bra}{\begin{array}}
\newcommand{\era}{\end{array}}
\newcommand{\beq}{\begin{equation}}
\newcommand{\eeq}{\end{equation}}
\newcommand{\beqar}{\begin{eqnarray}}
\newcommand{\eeqar}{\end{eqnarray}}

\def\BC{\bb C}
\def\_\BC{\bbi C}

\def\( {\left(}
\def\) {\right)}
\def\[ {\left[}
\def\] {\right]}
\def\no2 {{\textstyle{n\over 2}}}

\def\dag {{\dagger}}

\begin{document}
\thispagestyle{empty}
\begin{center}
\vspace{1.8cm}
 \renewcommand{\thefootnote}{\fnsymbol{footnote}}
 {\Large {\bf Amplified quantum battery via dynamical modulation}}\\


\vspace{1.5cm} {\bf Maryam Hadipour}$^{1}$, {\bf Negar Nikdel Yousefi}$^{2}$, {\bf Ali Mortezapour }$^{3}$ {\footnote { email: {\sf
mortezapour@guilan.ac.ir}}}, {\bf Amir Sharifi Miavaghi}$^{1}$ and {\bf Soroush Haseli}$^{1,4}$ {\footnote { email: {\sf
soroush.haseli@uut.ac.ir}}} 
\vspace{0.5cm}

$^{1}$ {\it  Faculty of Physics, Urmia University of Technology, Urmia, Iran}\\ [0.3em]
$^{1}$ {\it  Quantum Technologies Research Center (QTRC), Islamic Azad University, Science and Research Branch, Tehran, Iran.}\\ 
$^{3}$ {\it  Department of Physics, University of Guilan, Rasht 41335-1914, Iran}\\ 
$^{4}$ {\it  School of Nano Science, Institute for Research in Fundamental Sciences (IPM), P.O. 19395-5531, Tehran, Iran}\\ 
\end{center}


\baselineskip=18pt
\medskip
\vspace{3cm}
\begin{abstract}
We investigate the charging dynamics of a frequency-modulated quantum battery (QB) placed within a dissipative cavity environment. Our study focuses on the interaction of such a battery under both weak and strong coupling regimes, employing a model in which the quantum battery and charger are represented as frequency-modulated qubits indirectly coupled through a zero-temperature environment. It is demonstrated that both the modulation frequency and amplitude are crucial for optimizing the charging process and the ergotropy of the quantum battery.  Specifically, high-amplitude, low-frequency modulation significantly enhances charging performance and work extraction in the strong coupling regime. As an intriguing result, it is deduced that modulation at very low frequencies leads to the emergence of energy storage and work extraction in the weak coupling regime. Such a result can never be achieved without modulation in the weak coupling regime. These results highlight the importance of adjusting modulation parameters to optimize the performance of quantum batteries for real-world applications in quantum technologies.
\end{abstract}
\vspace{1cm}

\newpage
 \renewcommand{\thefootnote}{*}
\section{Introduction}\label{sec1}
Every quantum system is inherently open to interactions with its environment. These inevitable interactions can result in the loss of crucial quantum properties such as entanglement, coherence, and system energy disturbances  \cite{1}. The preservation of these quantum properties for efficient energy storage is a key area of study in the theory of open quantum systems. Therefore, it's vital to consider the dissipative effects of the environment during the charging process of quantum batteries (QBs). This understanding necessitates the study of quantum batteries within the framework of open quantum systems  \cite{2,3,4,5,6,7,7a,7b,7c,8,9,10,11,12,13,14,15,16,17,18,19,20,21,22,23,24,25,26,27,28,29,30,31,32,33,34,35,36,37,38,39,40,41,42,43,44,45,46,47,48,a1,a2,a3}. QBs are d-dimensional quantum systems with non-degenerate energy levels. These devices are used to storage of energy in quantum degrees of freedom for facilitating the transfer of energy from production to consumption hubs. A desirable quantum battery should possess two important features: firstly, it should be optimally charged within a short period of time, and secondly, its stored energy should be fully utilized at consumption centers, or in other words, the extraction of work form QBs should occur optimally. Hence, establishing the conditions to attain these two crucial characteristics for quantum batteries is of significant importance. Recent studies in the field of QBs have concentrated on scenarios of open systems, where the battery and charger in contact with the environment \cite{3,4,5,6,7,7a,8,9}. 

Motivated by these considerations, we aim to examine how frequency modulation can affect the dynamics of the charging process of QBs and work extraction from it in a model in which the charger-battery model is represented via a two-frequency-modulated qubit system. Generally, a quantum system experiences frequency modulation when its energy levels are changed by external driving. Frequency modulation in an atomic qubit can be performed by applying an external off-resonant field \cite{m20, 52}. On the other hand, frequency modulation is now possible in superconducting Josephson qubits (artificial atoms), which are the preferred building blocks of contemporary quantum computer prototypes \cite{m22}, thanks to recent experimental advancements in the fabrication and control of quantum circuit-QED devices \cite{53,m22,m24,m25,m26}. It was shown that frequency modulation of a qubit could induce sideband transitions \cite{49, m28}, modify its fluorescence spectrum\cite{51}, alter population dynamics \cite{m30,50,m32,m33,m34}, amplify non-Markovianity\cite{m35,m36}, quantum synchronization \cite{m39} and preserving quantum resources \cite{m36,m37}. Furthermore, external control of the frequency of the qubit has facilitated the identification of an accurate relationship between non-Markovianity and quantum speed limit time (QSLT) \cite{m38}.

Here, we focus on the scenario where the environment acts as a mediator between the QB and the charger. Indeed, this model allows for energy leakage from the battery into the environment, thereby facilitating a realistic  situation of spontaneous discharge in quantum batteries.  Our study adopts a model where a battery-charger system consists of two distinct frequency-modulated qubits interacting with a structured global environment. We consider both weak and strong coupling regimes in our model. This allows us to investigate how frequency modulation influences the charging process and work extraction across these regimes. The model we consider involves two distinct frequency-modulated qubits, the QB” and the charger,” interacting with a global structured environment. We analyze this interaction to understand the dynamics and behavior of the charging and work extraction of the QB. The findings of our study could potentially be applied in the design and optimization of quantum battery systems. We observe a significant disruptive effect of the modulation mechanism when the resonance condition is present, against the optimal charging and work extraction from the quantum battery in both strong and weak coupling regimes.
 Furthermore, it is demonstrated that under non-resonant conditions, both the charging and the work extraction process of the QB can be carried out more optimally.

The paper is structured as follows: Sec. \ref{Ergotropy} defines ergotropy and the work extraction mechanism from a quantum battery. Sec. \ref{model}, presents the model, detailing the system’s Hamiltonian and state evolution. In Sec. \ref{sec.RD}, we investigate the the charging performance of the quantum battery under different modulation conditions, analyzing the effects of frequency and amplitude on energy storage and work extraction in both weak and strong coupling regimes.  The main conclusions are represented in Sec. \ref{Conclusion}.

\section{Ergotropy}
\label{Ergotropy}
Let us consider a scenario where the quantum battery (QB) is thermally insulated, preventing heat exchange with its surroundings. In this scenario, the work extraction process is cyclic, implying that the system returns to its initial Hamiltonian by the end of the process. The following unitary transformation can characterize this process;
\begin{equation}
U(t)=\exp \left[ -i \int_0^t dt^\prime (H_B+V(t^{\prime}))\right].
\end{equation}

Where $H_B$ is the Hamiltonian of the QB. $V(t^{\prime})$ represents the time-dependent fields that extract energy from the quantum battery. In this paper, we have normalized the constant ($\hbar$) to 1 ($\hbar = 1$). Given that the process is cyclic, the field is zero at both the initial and final moments, i.e., $V(0) = V(t)=0$. The work that can be extracted during this cyclic process is as follows:

\begin{equation}
W(\rho_B)=Tr(H_B \rho_B)-Tr(H_B U \rho_B U^\dag),
\end{equation}
in which $\rho_B$ is the density matrix of the QB. By appropriately selecting $ V(t^{\prime})$, any unitary transformation $U$ can be realized. Consequently, the maximum work that can be extracted from a quantum battery in a cyclic unitary process referred to as ergotropy, is determined by
\begin{equation}
\mathcal{W}=Tr(H_B \rho_B) - \min_U (H_B U \rho_B U^\dag),
\end{equation}
The minimization is computed over the ensemble of all attainable unitary transformations. It has been demonstrated that for any specified state $\rho_B$, there exists a singular state that optimizes the aforementioned relation. This state is referred to as the passive state \cite{54}. So, the maximum amount of extractable work can be expressed as
\begin{equation}\label{ergo}
\mathcal{W}=Tr(H_B \rho_B)-Tr(H_B \sigma_B),
\end{equation}
where $\sigma_B$ is the passive state associated with $\rho_B$. The passive state $\sigma_B$ is characterized by a population distribution that decreases or remains constant concerning its Hamiltonian $H_B$, and it commutes with $H_B$, satisfying $[H_B, \sigma_B] = 0$. The spectral decomposition of the density matrix $\rho_B$ and its associated Hamiltonian $H_B$ can be expressed in the following form
\begin{eqnarray}\label{spec}
\rho_B&=&\sum_n r_n \vert r_n \rangle \langle r_n \vert, \quad r_1 \geq r_2 \geq ... \geq r_n \\  
H_B&=&\sum_m \varepsilon_{m} \vert \varepsilon_m \rangle \langle \varepsilon_m \vert  \quad \varepsilon_1 \leq \varepsilon_2 \leq ... \leq \varepsilon_m, \nonumber
\end{eqnarray}
in which $r_n$($\vert r_n \rangle$) and $\varepsilon_m$($\vert \varepsilon_m \rangle$ ) are the eigenvalues (eigenstate) of the density matrix and Hamiltonian, respectively. Hence, the passive state $\sigma_B$ can be written as 
\begin{equation}\label{pass}
\sigma_B= U \rho_B U^{\dag}=\sum_i r_n \vert \varepsilon_n \rangle \langle \varepsilon_n \vert.
\end{equation}
In general,  by substituting Eq.\ref{spec} and Eq. \ref{pass} into Eq.\ref{ergo}, the ergotropy can be obtained as
\begin{equation}
\mathcal{W}=\sum_{mn} r_n \varepsilon_m \left( \vert \langle r_n \vert \varepsilon_m \rangle \vert^{2} - \delta_{m,n} \right),
\end{equation}
where $\delta_{m,n}$ is the Kronecker delta function.
\section{Model}
\label{model}
\begin{figure}
    \centering
    \includegraphics[width =0.5 \linewidth]{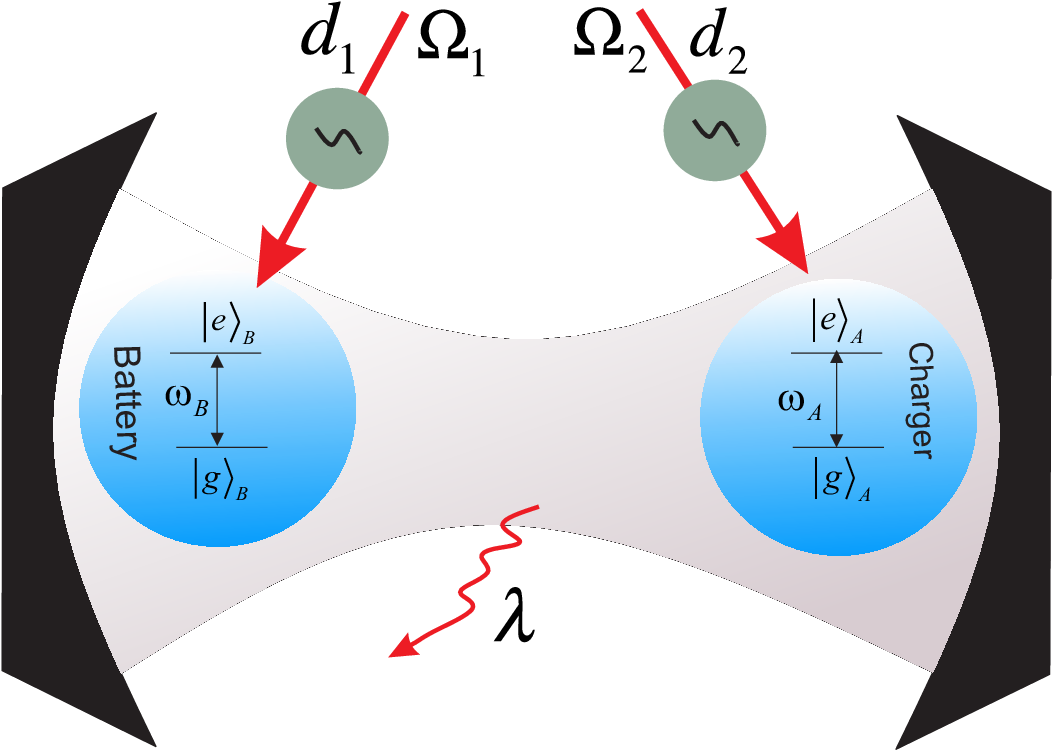}
    \centering
    \caption{Schematic representation of the model in which QB and charger (two frequency-modulated qubits) are embedded in a high-Q cavity}
    \label{Fig1}
  \end{figure}

The model considered in this work consists of the QB (qubit $B$) and the charger (qubit $A$). Each qubit has ground and excited states, represented as $\vert g \rangle$ and $\vert e \rangle$, respectively. Both qubits are embedded in a high-$Q$ cavity with quantized modes and are influenced by a zero-temperature environment, which acts as a mediator between the QB and the charger. The QB and charger are modulated sinusoidally by an external driving field. A schematic representation of the model is illustrated in Fig. \ref{Fig1}.  
The total Hamiltonian which describes the system consist of QB, charger and environment  in the dipole and rotating wave approximations is given by 
\begin{equation}\label{totalh}
H=H_0 + \mathcal{F}(t) H_I
\end{equation}
$H_0=H^{m}_A+H^{m}_B+H_E$, is the Hamiltonian of the total system,  where the first two terms and second term are the Hamiltonian of the modulated two-qubit system and environment respectively, they  are given by
\begin{eqnarray}
H^{m}_{j}&=& \sum_{j} \frac{1}{2}[\omega_j + d_j \cos(\Omega_j t)] \sigma_z^j, \quad j \in \lbrace A,B \rbrace  \\
H_E &=& \sum_k \omega_k a^{\dag}_k a_k. \nonumber
\end{eqnarray}
in the above expression, $\sigma_z^{j}$ represents the $z$-component of the Pauli operator for the $j$th qubit, which has a transition frequency $\omega_j$. The variables $d_j$ and $\Omega_j$ denote the amplitude and frequency of the modulation applied to the $j$th qubit, respectively. $\omega_k$ is the frequency of the quantized modes of the cavity, while $a_k$ and $a^{\dag}_k$ are the annihilation and creation operators for the $k$th mode of the cavity. In Eq. \ref{totalh}, $H_I$ describes the interaction between the two qubits and the reservoir, which is expressed as follows:
\begin{equation}
\begin{aligned}
& H_I=H_{A E}+H_{B E}=\sum_k g_k \mu_1\left(\sigma_{\mathrm{A}}^{+} a_k+\sigma_{\mathrm{A}}^{-} a_k^{\dagger}\right) \\
& +\sum_k g_k \mu_2\left(\sigma_{\mathrm{B}}^{+} a_k+\sigma_{\mathrm{B}}^{-} a_k^{\dagger}\right),
\end{aligned}
\end{equation}
where $\sigma_j^+$ and $\sigma_j^{-}$ are the raising and lowering operator for the $j$th qubit. $g_k \mu_i$ is the coupling strength between qubits and $k$th mode of the cavity, where $\mu_i$ is a dimensionless real parameter. The relative interaction strength and the collective coupling are $r_i = \mu_i / \mu_T$ and $\mu_T=(\mu_1^2+\mu_2^2)^{1/2}$, respectively. The dimensionless function $\mathcal{F}(t)$ in Eq. \ref{totalh} is given by
\begin{equation}
\mathcal{F}(t)= \begin{cases} 1 & t \in [0,\tau ), \\ 0 & \text {elsewhere,} \end{cases}
\end{equation}
which is used to switch interactions on or off, and $\tau$ is the charging time of the QB. It is assumed that for $t<0$, charger and QB do not interact with reservoir, while when $t=0$, charger and battery start their interaction with the reservoir by switching on $H_{AE}$ and $H_{BE}$, respectively. In this scenario, the charger and QB do not interact with each other. Given that $[H_B+H_E, H_{BE}] \neq 0$, and assuming the environment is in its ground state at $t=0$, it is plausible that the final energy of the quantum battery (QB) is not solely derived from the charger $A$. Furthermore, the thermodynamic work involved in modulating the interactions at specific switching times may contribute to the final energy of the QB. Therefore, within the time interval $[0,\tau)$, a fraction of the energy from the charger is transferred to the quantum battery (QB) through the interaction facilitated by the environment. At the final time, $\tau$, the charger and the quantum battery (QB) are once again isolated by severing the connections between $A$ and $E$ and between $B$ and $E$. It is important to note that in our model, there is no direct interaction between the charger and the quantum battery (QB). Consequently, we investigate a wireless charging mechanism for the QB, wherein the environment acts as an intermediary in the charging process. Subsequently, we will analyze the process of charging and work extraction in the considered model based on the introduced Hamiltonian. The local unitary transformation of the following form will be considered as 
\begin{equation}
\begin{aligned}
\hat{U} & =\vec{T} \exp \left[-i \int_0^t\left(\hat{H}_{\mathrm{S}}(\tau)+\hat{H}_{\mathrm{E}}(\tau)\right) d \tau\right] \\
& =\exp \left[-i\left\{\frac{1}{2}\left[\omega_B t+\left(d_B / \Omega_B\right) \sin \left(\Omega_B t\right)\right] \hat{\sigma}_z^{(B)}\right.\right. \\
& \left.\left.+\frac{1}{2}\left[\omega_A t+\left(d_A / \Omega_A\right) \sin \left(\Omega_A t\right)\right] \hat{\sigma}_z^{(A)}+\sum_k \omega_k \hat{a}_k^{\dagger} \hat{a}_k t\right\}\right]
\end{aligned}
\end{equation}
where $\vec{T}$ is the time-ordering operator. So, the effective Hamiltonian can be obtained as 
\begin{equation}\label{effh}
H_{eff}=\sum_{j,k} \left( \mu_j \sigma_j^+ g_k a_k e^{-i(\omega_k - \omega_j)t}e^{i(d_j/\Omega_j)\sin \Omega_j t}+ H.c. \right).
\end{equation}
To obtain the above effective Hamiltonian, we have used $H_{eff}=U^{\dag}HU + i (\partial U^{\dag}/\partial t) U$. Utilizing the Jacobi-Anger expansion, the exponential terms in Eq.\ref{effh} can be expressed as
\begin{equation}
e^{\pm i (d_j/\Omega_j)\sin \Omega_j t}= J_0\left( \frac{d_j}{\Omega_j}\right) +2 \sum_{n=1}^{\infty}(\pm i)^n J_n\left( \frac{d_j}{\Omega_j}\right) \cos(n \Omega_j t),
\end{equation}
in which $J_n \left( \frac{d_j}{\Omega_j} \right)$ is the $n$-th Bessel function of the first kind. Let us examine the initial state, which is expressed in the following form
\begin{equation}\label{inits}
\left|\psi_0\right\rangle=\left(c_{01}|e_A, g_B\rangle+c_{02}|g_A, e_B\rangle\right)|\mathbf{0}\rangle_R
\end{equation}
where $\vert \mathbf{0}\rangle_R$ is the multi-mode vacuum state. Eq.\ref{inits}  indicates no excitations in the cavity modes, and the qubits are entangled.  
The state of the system at time $t$ can be derived as 
\begin{equation}\label{statet}
\begin{aligned}
|\psi(t)\rangle= & c_1(t) |e, g\rangle|\mathbf{0}\rangle_R+c_2(t) |g, e\rangle|\mathbf{0}\rangle_R \\
& +\sum_k c_k(t) |g, g\rangle\left|\mathbf{1}_k\right\rangle,
\end{aligned}
\end{equation}
Here, $\vert \mathbf{1}_k \rangle$ denotes the state of the cavity with a single excitation in the $k$-th mode. By substituting Equation \ref{statet} into the Schr$\ddot{o}$dinger equation $i \frac{d}{dt} \vert \psi(t) \rangle = H_{eff} \vert \psi(t) \rangle$ and performing the required algebraic manipulations, we derive the following integro-differential equations. 
\begin{equation}\label{integro1}
\begin{aligned}
\dot{C}_1(t)=- & \sum_k \int_0^t\left|g_k\right|^2 \mathrm{~d} t^{\prime}\left(\mu_1^2 e^{i d_B / \Omega_B\left(\sin \left(\Omega_B t\right)-\sin \left(\Omega_B t^{\prime}\right)\right)} e^{i \delta_k^{(B)}\left(t-t^{\prime}\right)} C_1\left(t^{\prime}\right)\right. \\
& \left.+\mu_1 \mu_2 e^{i d_B / \Omega_B \sin \left(\Omega_B t\right)} e^{-i d_A / \Omega_A \sin \left(\Omega_A t^{\prime}\right)} e^{i \delta_k^{(B)} t} e^{-i \delta_k^{(A)} t^{\prime}} C_2\left(t^{\prime}\right)\right), \\
\dot{C}_2(t)=- & \sum_k \int_0^t\left|g_k\right|^2 \mathrm{~d} t^{\prime}\left(\mu_2^2 e^{i d_A / \Omega_A\left(\sin \left(\Omega_A t\right)-\sin \left(\Omega_A t^{\prime}\right)\right)} e^{i \delta_k^{(A)}\left(t-t^{\prime}\right)} C_2\left(t^{\prime}\right)\right. \\
& \left.+\mu_1 \mu_2 e^{-i d_B / \Omega_B \sin \left(\Omega_B t^{\prime}\right)} e^{i d_A / \Omega_A \sin \left(\Omega_A t\right)} e^{-i \delta_k^{(B)} t^{\prime}} e^{i \delta_k^{(A)} t} C_1\left(t^{\prime}\right)\right) .
\end{aligned}
\end{equation}
where $\delta_k^{(j)}=\omega_k - \omega_j$. When considering the continuous spectrum of the environment frequencies, we transition from summation over modes to integration using the following relation: $\sum_k \vert g_k \vert^2 \rightarrow \int d\omega J(\omega)$, where $J(\omega)=W^2 \lambda / \pi [(\omega - \omega_c)+ \lambda^2]$ is the Lorentzian spectral density with $\omega_c$ is the fundamental frequency of the cavity. In this context,  $W$ is proportional to the vacuum Rabi frequency ($\mathcal{R} = \alpha_T W$), and $\lambda$ represents the cavity losses. By defining the correlation function as $f(t,t^{\prime})=\int d \omega J(\omega) e^{i(\omega_c-\omega)(t-t^{\prime})}$, Eq.\ref{integro1} transform into the following form
\begin{equation}\label{integrof}
\begin{aligned}
\dot{C}_1(t) & =-\int_0^t \mathrm{~d} t^{\prime} f\left(t, t^{\prime}\right)\left(\mu_1^2 e^{i d_B / \Omega_B\left(\sin \left(\Omega_B t\right)-\sin \left(\Omega_B t^{\prime}\right)\right)} C_1\left(t^{\prime}\right)\right. \\
& \left.+\mu_1 \mu_2 e^{i d_B / \Omega_B \sin \left(\Omega_B t\right)} e^{-i d_A / \Omega_A \sin \left(\Omega_A t^{\prime}\right)} e^{-i \delta_{AB} t^{\prime}} C_2\left(t^{\prime}\right)\right) e^{i \delta_B\left(t-t^{\prime}\right)}, \\
\dot{C}_2(t) & =-\int_0^t \mathrm{~d} t^{\prime} f\left(t, t^{\prime}\right)\left(\mu_1 \mu_2 e^{-i d_B / \Omega_B \sin \left(\Omega_B t^{\prime}\right)} e^{i d_A / \Omega_A \sin \left(\Omega_A t\right)} e^{i \delta_{AB} t^{\prime}} C_1\left(t^{\prime}\right)\right. \\
& \left.+\mu_2^2 e^{i d_A / \Omega_A\left(\sin \left(\Omega_A t\right)-\sin \left(\Omega_A t^{\prime}\right)\right)} C_2\left(t^{\prime}\right)\right) e^{i \delta_A\left(t-t^{\prime}\right)},
\end{aligned}
\end{equation}
where $\delta_j=\omega_j - \omega_c$ and $\delta_{AB}=\omega_A-\omega_B$. Consider the special case where the two qubits are identical, such that $\omega_B=\omega_A \equiv \omega_0$, $\Omega_B = \Omega_A \equiv \Omega$ and $d_B=d_A \equiv d$. So, we have $\delta_A=\delta_B \equiv \delta$ and $\delta_{AB}=0$. Considering these situations Eq. \ref{integrof} reduces to 
\begin{equation}\label{integrop}
\begin{aligned}
& \dot{C}_1(t)=-\int_0^t \mathrm{~d} t^{\prime} F\left(t, t^{\prime}\right)\left(\mu_1^2 C_1\left(t^{\prime}\right)+\mu_1 \mu_2 C_2\left(t^{\prime}\right)\right), \\
& \dot{C}_2(t)=-\int_0^t \mathrm{~d} t^{\prime} F\left(t, t^{\prime}\right)\left(\mu_2^2 C_2\left(t^{\prime}\right)+\mu_1 \mu_2 C_1\left(t^{\prime}\right)\right),
\end{aligned}
\end{equation}
where $F(t,t^{\prime})=W^2 e^{(-\lambda + i \delta)(t-t^{\prime})}e^{i d/\Omega (\sin(\Omega t)-\sin(\Omega t^{\prime}))}$. 

We set $\dot{C}_i = 0$ in Eq.\ref{integrop} to achieve a stable entangled state. This produces the following long-lived, decoherence-free state
\begin{equation}
\vert \psi_{-} \rangle = r_2 \vert e_A, g_B \rangle - r_1 \vert g_B, e_A \rangle. 
\end{equation}
The above state experiences no decoherence or change over time. Since $|\psi_{-}\rangle$ remains static, the only observable time evolution is that of its orthogonal state $\vert \psi_+ \rangle = r_1 \vert e_a, g_B \rangle + r_2 \vert g_A, e_B \rangle$. The survival amplitude of the above state can be described as $\mathcal{E}(t)= \langle \psi_+ \vert \psi_+(t) \rangle$. The survival amplitude is governed by the following integro-differential equation of motion
\begin{equation}
\dot{\mathcal{E}}(t)= \mu_T^{2} \int_0^t F(t,t^{\prime}) \mathcal{E}(t^{\prime}) dt^{\prime}.
\end{equation}
so, the time-dependent probability amplitude in Eq.\ref{statet} can be derived as follows
\begin{equation}
\begin{aligned}
& c_1(t)=\left[r_2^2+r_1^2 \mathcal{E}(t)\right] c_{01}-r_1 r_2[1-\mathcal{E}(t)] c_{02}, \\
& c_2(t)=-r_1 r_2[1-\mathcal{E}(t)] c_{01}+\left[r_1^2+r_2^2 \mathcal{E}(t)\right] c_{02} .
\end{aligned}
\end{equation}

\section*{Results and discussion}
\label{sec.RD}
This section provides a comprehensive analysis of each step involved in the quantum battery (QB) charging process within the specified model. By performing a partial trace over each subsystem $A$ and $B$ in Eq. \ref{statet}, we can obtain the reduced time-dependent density matrices corresponding to the QB and charger at the specific time $t = \tau$,
\begin{equation}
\begin{aligned}
& \rho_B(\tau)=\vert c_2(\tau) \vert^{2} \vert e \rangle_B \langle e \vert \left( 1-\vert c_2(\tau) \vert^{2} \right) \vert g \rangle _B \langle g \vert,  \\
& \rho_A(\tau)=\vert c_1(\tau) \vert^{2} \vert e \rangle_A \langle e \vert \left( 1-\vert c_1(\tau) \vert^{2} \right) \vert g \rangle _A \langle g \vert,
\end{aligned}
\end{equation}
To explore the relation between the energy of the QB and the charger, it is beneficial to analyze their respective energy variations. The variation in the internal energy of the QB during the charging process can be expressed as follows:
\begin{equation}
\Delta E_B =Tr \left[ H_B \rho_B(\tau)\right]-Tr\left[ H_B \rho_B(0)\right], 
\end{equation} 
where $H_B=(\omega_0/2) \sigma_z$ is the free Hamiltonian of the QB and $\rho_B(\tau)$ is the state of the QB at time $\tau$. So, the changes in the internal energy of the QB $\Delta E_B$ can be derived as 
\begin{equation}
\Delta E_B = \omega_0 \left( \vert c_2(\tau) \vert^2 - \vert c_2(0) \vert^2 \right).
\end{equation}
The maximum work that can be extracted from a quantum battery (QB) at the end of the charging process, when subjected to cyclic unitary operations, is defined as 
\begin{equation}
\mathcal{W}= \mathcal{W}_{max}\left( 2 \left| c_2(\tau) \right|^2 - 1\right)\Theta\left( \left| c_2(\tau) \right|^2 - \frac{1}{2}\right),  
\end{equation}
where $\Theta(x)$ is the Heaviside function and $\mathcal{W}_{max}=\omega_0$.

Our initial assumption is that the quantum battery (QB) is entirely devoid of energy, while the charger contains significantly more internal energy than the QB. This scenario arises when the state of the composite quantum system Encompassing the quantum battery (QB), the charger, and the environment can be described as follows:

\begin{equation}
\vert \psi_o \rangle = \vert e \rangle_A \vert g \rangle_B \otimes \vert \textbf{0} \rangle_R, \quad ( c_{01}=1, c_{02}=0). 
\end{equation}

\begin{figure}
	\centering
	\includegraphics[width =0.6 \linewidth]{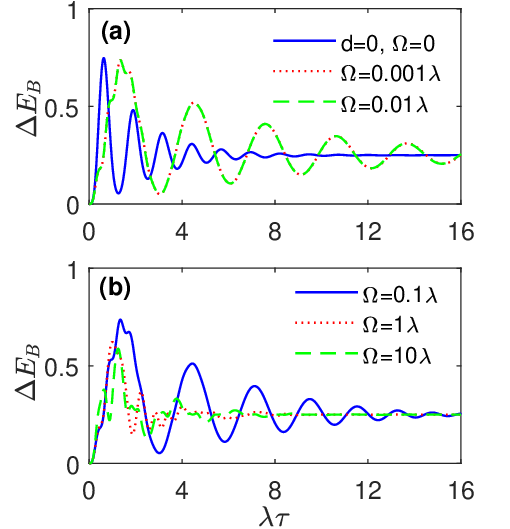}
	\centering
	\caption{The energy change of QB ($\Delta E_B$) as a function of scaled time $\lambda \tau$ for different values of the modulation frequency $\Omega$ in the strong coupling regime with $R=5 \lambda$. The values of other parameters are: $d=10\lambda$ and  $r_1=r_2=1/\sqrt{2}$. Solid-blue line in panel (a) corresponds to the situation in which frequency modulation is off. }
	\label{Fig2}
\end{figure}

We begin the quantitative analysis by studying the time evolution of the energy change of the QB ($\Delta E_B$) under a strong coupling regime ($R=5 \lambda$), plotted in Fig. \ref{Fig2} for different values of the modulation frequency ($\Omega$) and fixed modulation amplitude $d=10\lambda$. This evolution is compared to the situation without a modulation process ($d=0$, $\Omega=0$). In this case, fluctuations disappear around time $ \lambda \tau \sim 10$. Frequent modulation improves the energy dynamics of the quantum battery for $\Omega \leq 1$. Namely, low-frequency modulation delays the process of reaching a steady state. On the other hand, the energy variation of the quantum battery (QB) decreases as the modulation frequency increases (Fig.\ref{Fig2}(b)). The mechanism behind this is related to the fact that when the modulation frequency becomes large, the system dynamics lose their ability to synchronize effectively with the driving modulation. Instead, the rapid oscillations induced by this high-frequency modulation tend to cancel each other over time. As a result, the energy exchange between the QB and its surrounding environment is significantly reduced.  This effect, known as "dynamical decoupling," occurs because the system struggles to absorb or transfer energy when the modulation frequency goes beyond a certain limit. As a result, the energy change in the QB decreases as the modulation frequency increases in this high-frequency range.

\begin{figure}
    \centering
    \includegraphics[width =0.6 \linewidth]{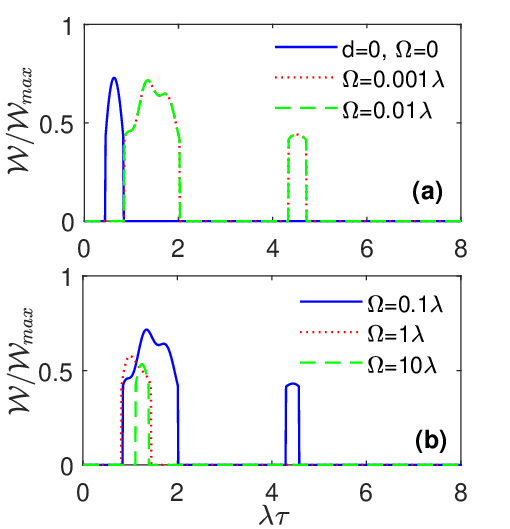}
    \centering
    \caption{The dynamics behaviour of $\mathcal{W}/\mathcal{W}_{\max}$ as a function of scaled time $\lambda \tau$. Other parameters are taken as
    	in Fig. \ref{Fig2} }
    \label{Fig3}
  \end{figure}

Fig. \ref{Fig3} displays the ergotropy versus the scaled time parameter $\lambda \tau$ for various modulation frequencies. Other parameters are taken as in Fig. \ref{Fig2}. It is seen that the ergotropy can be significantly affected by frequency modulation. For $\Omega \leq 1$, both the amount of work that can be extracted from the system and the time interval for extracting this work increase compared to the off-modulation scenario. Conversely, as omega increases after $\Omega=1$, the amount of work that can be extracted and the required time interval decrease.
This results in a disruption of the system's coherence, and the quantum correlations that are necessary for efficient work extraction are weakened. As a consequence, the system is unable to reach a state of optimal energy alignment for work extraction, leading to a decrease or destruction of the ergotropy.

\begin{figure}
    \centering
    \includegraphics[width =.8 \linewidth]{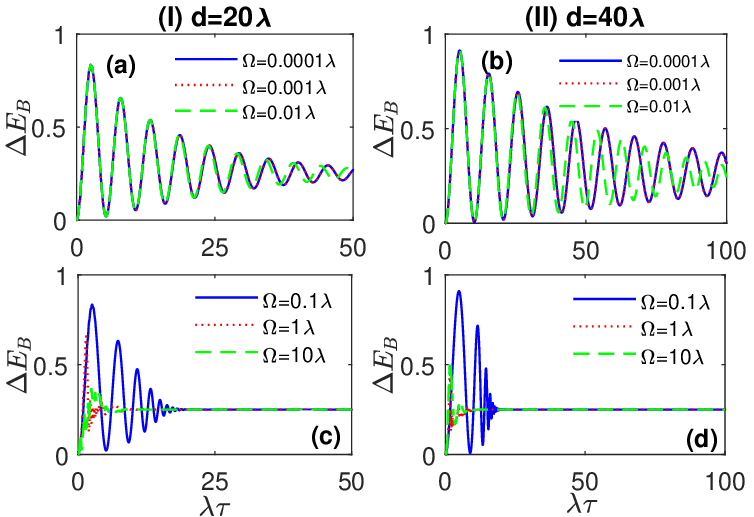}
    \centering
    \caption{The energy change of QB ($\Delta E_B$)  as a function of scaled time $\lambda \tau$ for different values of the modulation frequency $\Omega$ in the strong coupling regime with $R=5 \lambda$. The values of other parameters are: $r_1=r_2=1/\sqrt{2}$, $d=20$ for column (I) and $d=40$ for column (II).}
    \label{Fig4}
  \end{figure}
  
\begin{figure}
    \centering
    \includegraphics[width =0.8 \linewidth]{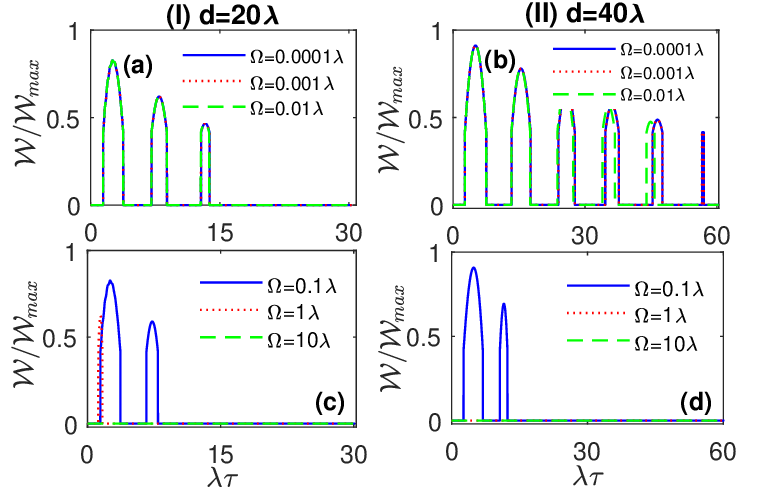}
    \centering
    \caption{The dynamics behaviour of $\mathcal{W}/\mathcal{W}_{\max}$ as a function of scaled time $\lambda \tau$ for different values of the modulation frequency $\Omega$. Other parameters are taken as in Fig. \ref{Fig4}. }
    \label{Fig5}
  \end{figure}

We now aim to examine how variations in modulation amplitude impact the time evolution of $\Delta E_B$ and ergotropy, as illustrated in Fig. \ref{Fig4} and Fig. \ref{Fig5}, respectively. Both columns of Fig. \ref{Fig4} indicate that a decrease in modulation frequency substantially prolongs the duration of $\Delta E_B$ oscillations. This finding aligns with the results presented in Fig. \ref{Fig2}. Specifically, an increase in modulation frequency acts as a disruptive factor, further diminishing energy variations. Furthermore, by comparing the curves in the two columns of Fig. \ref{Fig4} with those depicted in Fig. \ref{Fig2}, we observe that for a fixed modulation frequency, an increase in modulation amplitude not only lengthens the duration of the oscillations but also enhances their amplitude.

In Fig. \ref{Fig5}, it is clear that an increase in modulation amplitude considerably enhances the ergotropy of the system. A comparison of the blue solid line in Fig. 3 with the curves shown in Fig. \ref{Fig5} highlights the crucial role of modulation in boosting ergotropy. In such a way, increasing the modulation amplitude while simultaneously reducing the modulation frequency leads to a substantial rise in the amount of work that can be extracted and extending the time intervals available for work extraction.

  \begin{figure}
    \centering
    \includegraphics[width =0.8 \linewidth]{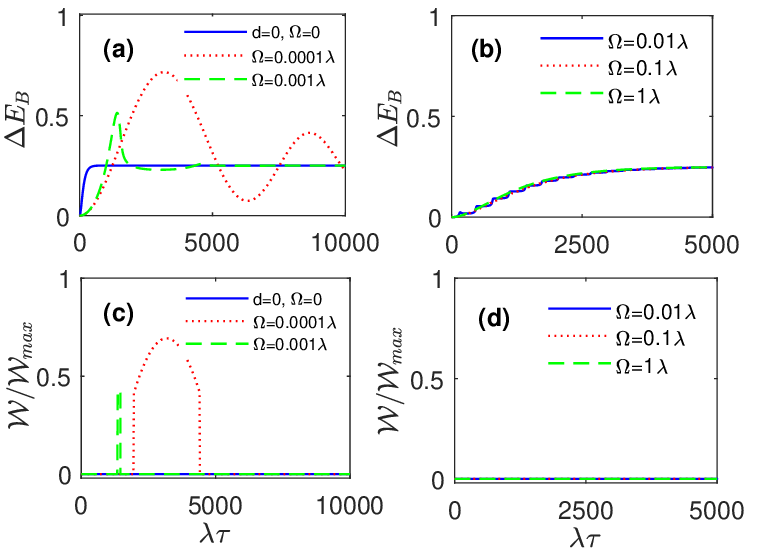}
    \centering
    \caption{The dynamics behaviour of $\Delta E_B$ [(a) , (b)] and $\mathcal{W}/\mathcal{W}_{\max}$ [(c) , (d)] as functions of scaled time $\lambda \tau$  for different values of the modulation frequency $\Omega$ in the weak coupling regime with $R=0.1 \lambda$. The values of other parameters are: $r_1=r_2=1/\sqrt{2}$ and $d= 10 \lambda$. Solid-blue line in panels ($a$) and ($b$) corresponds to the situation in which frequency modulation is off.}
    \label{Fig6}
  \end{figure}
 
In Fig. \ref{Fig6}, we turn to the weak coupling regime, where the time evolution of $\Delta E_B$ [(a), (b)] and $\mathcal{W}/\mathcal{W}_{\max}$ [(c), (d)] is plotted for different modulation frequencies with $R=0.1$. Here, the solid-blue line in Fig. \ref{Fig6} (a) and Fig. \ref{Fig6}(c) denotes the scenario in which modulation is turned off ($\Omega= 0$ and $d = 0$ ). It is seen that, in the absence of modulation, the energy change of QB reaches a steady-state value from a starting point of zero without any oscillation, resulting in zero ergotropy. However, when modulation is applied, especially at low frequencies, oscillations emerge in the dynamic behaviour of $\Delta E_B$, enabling work extraction from the system. Therefore, oscillations of $\Delta E_B$ increase as the modulation frequency decreases, leading to a greater capacity for work extraction. Conversely, as the modulation frequency rises, the oscillations of $\Delta E_B$ and the associated ergotropy diminish. Fig. \ref{Fig6} presents an intriguing result, demonstrating that work can be extracted from the system via modulation, even in the weak coupling regime. However, this phenomenon occurs this phenomenon occurs at very low frequencies.

Finally, it is important to note that our calculations indicate that adjusting the ratio of the amplitude and frequency of modulation at the zeros of the Bessel function of various orders—by changing the modulation frequency—will not yield better results than the situation in which the modulation is off.

\section*{Conclusion}
\label{Conclusion}
This work investigated the charging performance of a frequency-modulated quantum battery (QB) embedded in a dissipative cavity environment. By modeling the system as two frequency-modulated qubits coupled to a common zero-temperature reservoir, we explored how modulation parameters influence the charging process and ergotropy under both weak and strong coupling regimes. Our results reveal that frequency modulation plays a crucial role in optimizing the performance of the quantum battery. In the strong coupling regime, we found that high-amplitude, low-frequency modulation significantly enhances both the charging efficiency and the extractable work. However, the charging performance degrades for higher modulation frequencies due to disruptive effects on energy transfer and coherence. A particularly noteworthy result emerged in the weak coupling regime: while work extraction is typically infeasible in this regime, we demonstrated that applying very low-frequency modulation enables energy storage and work extraction. This effect is otherwise impossible without modulation. This finding is of fundamental importance, as it suggests that even under weak coupling conditions, the strategic use of frequency modulation can transform an inefficient quantum battery into a functional and extractable energy resource. Overall, our findings underscore the importance of modulation as a powerful tool for enhancing quantum battery performance. By carefully tuning the modulation parameters, one can significantly improve charging efficiency and ergotropy, even in regimes where work extraction was previously considered unachievable. These insights pave the way for practical implementations of quantum batteries in real-world quantum technologies, offering a new avenue for efficient energy storage and management in quantum systems.

\section*{Author contributions statement}
Maryam Hadipour,  Negar Nikdel Yousefi, Ali Mortezapour, Amir Sharifi Miavaghi and Soroush Haseli all contributed to the development and completion of the idea, performing the calculations, analyzing the results, discussions and writing the manuscript.
\section*{Data availability}
No data were created or analyzed in this study.
\section*{Disclosures}
The authors declare that they have no known competing financial interests.

\section*{ORCID iDs}
Maryam Hadipour \href{https://orcid.org/0000-0002-6573-9960}{https://orcid.org/0000-0002-6573-9960}\\
Ali Mortezapour \href{https://orcid.org/0000-0002-1709-2509}{https://orcid.org/0000-0002-1709-2509}\\
Soroush Haseli \href{https://orcid.org/0000-0003-1031-4815}{https://orcid.org/0000-0003-1031-4815}


\end{document}